\documentclass[12pt,preprint]{aastex}   
\usepackage{emulateapj5,epsfig}   
\usepackage{apjfonts}

\newcommand{\feh}{\hbox{$ [\mathrm{Fe}/\mathrm{H}]$}}   
   
\newcommand{\ea}{\hbox{et al.}}   
\newcommand{\wcen}{\hbox{$ \omega\,$Cen}}   
   
\slugcomment{to appear in ApJ Letters}   
   
\shorttitle{The Subgiant Mass Age of \wcen}   
\shortauthors{Chaboyer \& Krauss} 
     
\begin{document}      
\title{Theoretical Uncertainties in the Subgiant--Mass Age relation and 
the absolute age of \wcen}      
\author{Brian Chaboyer\altaffilmark{1}}
\affil{Department of Physics and Astronomy, Dartmouth College 6127 Wilder
Lab, Hanover, NH 03755}   
\email{Brian.Chaboyer@Dartmouth.edu}   
\and    
\author{Lawrence M.\ Krauss}   
\affil{Departments of Physics and Astronomy, Case Western Reserve  
University, 10900 Euclid Ave. Cleveland OH 44106-7079}   
\email{krauss@theory1.phys.cwru.edu}   

\altaffiltext{1}{Visiting Scholar, Astronomy Unit, Queen Mary,
University of London, Mile End Road, London E1 4NS, UK}

\begin{abstract}   
The theoretical uncertainties in the calibration of the relationship   
between the subgiant mass and age in metal-poor stars are investigated   
using a Monte Carlo approach.  Assuming that the mass and iron   
abundance of a subgiant star are known exactly, uncertainties in the   
input physics used to construct stellar evolution models and   
isochrones lead to a Gaussian 1-$\sigma$ uncertainty of $\pm 2.9\%$ in   
the derived ages.  The theoretical error budget is dominated by the   
uncertainties in the calculated opacities.   
   
Observations of detached double lined eclipsing binary OGLEGC-17 in   
the globular cluster \wcen\ have found that the primary is on the   
subgiant branch with a mass of $\mathrm{M} = 0.809\pm   
0.012\,\mathrm{M}_{\odot}$ and $\feh = -2.29\pm 0.15$ \citep{kaluzny}.   
Combining the theoretical uncertainties with the observational errors   
leads to an age for OGLEGC-17 of $11.10\pm 0.67\,$Gyr.  The one-sided,   
95\% lower limit to the age of OGLEGC-17 is 10.06 Gyr, while the   
one-sided, 95\% upper limit is 12.27 Gyr.    
   
\end{abstract}   
   
\section{Introduction}   
Traditionally, absolute globular cluster (GC) ages have been determined
using the absolute magnitude of the main sequence turn-off (TO), or
subgiant branch (SGB), as this minimizes the theoretical uncertainties
associated with stellar evolution models \citep[e.g.\
][]{renz91,mvsgb}.  This age determination method requires that the
distance to the GC be known.  There is considerable
uncertainty regarding the distance scale to GCs, and
this translates into a significant uncertainty in the absolute age
estimates of GC \citep{mc3}. To avoid this error
\cite{pacz} has advocated the use of detached eclipsing double line
spectroscopic binaries to determine the age of GCs.  In
these binary systems, it is possible to determine the mass of the
individual stars.  These mass estimates are derived in a fundamental
manner, and are likely to be free from systematic errors \citep{pacz}.
If one of the members of the binary is at the TO, or on the SGB then
the age of the cluster may be determined from the TO/SGB mass-age
relation.
   
In principle the relation between the TO/SGB mass and age is robust
prediction of stellar evolution theory -- it simply depends on the
amount of hydrogen fuel available for nuclear burning in the core of
the star and the luminosity of the star during its main sequence
lifetime.  Thus, the TO/SGB mass-age relation should be insensitive
to the details of what occurs near the surface of stars and will not
depend on the treatment of convection for the low mass stars in
GCs \citep{pacz}.  For these reasons, one might expect
that ages derived from the masses of TO/SGB stars will be relatively
insensitive to various significant uncertainties that might otherwise
be important in stellar structure calculations.  
   
This paper will explore how the uncertainties in stellar   
structure and evolution calculations (\S \ref{uncertain}) translate   
into errors in ages derived from SGB masses in GCs   
(\S \ref{monte}).  This work is motivated by the high precision   
mass estimate for the detached eclipsing double line spectroscopic   
binary OGLEGC-17 in \wcen\ by \cite{kaluzny}.  
The primary in OGLEGC-17 is on SGB \citep{thompson}.   
The age of this star is derived in \S \ref{age}, and this paper
concludes with a general discussion of the implications of this age
determination in \S \ref{universe}.
   
\section{Uncertainties in Stellar Evolution Models \label{uncertain}}   
The basic equations of stellar structure are simple -- hydrostatic
equilibrium, conservation of mass and energy, and an equation for
energy transfer.  However, the solution of these equations requires a
considerable amount of additional information -- composition of the
star must be specified, one needs to know opacities, nuclear reaction
rates, surface boundary conditions, etc.  There are uncertainties
associated with all of these, and these uncertainties in the input
physics lead to uncertainties in the calculated structure and
evolution of a star.  Furthermore, there are uncertainties associated
with the modeling of convection in stars, and indeed with the
inclusion of additional physical processes such as diffusion.  Given
that the equations of structure must be solved numerically, it is
easiest to evaluate the uncertainties associated with stellar
structure and evolution calculations using a Monte Carlo (MC) procedure
\cite[]{mc1}. Once the distribution of each input parameter is
specified, one randomly selects a specific value for each of the input
parameters and constructs stellar evolution models for a variety of
masses.  These stellar evolution models are then used to construct an
isochrone which can be used to derive the age of a GC. This procedure
is repeated numerous times and the result is a set of isochrones
which can be used to determine the error associated with stellar age
estimates.
   
Full details on our choice of parameters are in our previous papers
\citep{mc1,mc2,mc3}.  In brief, the following input parameter
distributions were used: mixing length $1.85\pm 0.25$; helium
diffusion coefficients multiplied by 0.2 -- 0.8 (flat distribution);
high temperature ($T > 10^4\,$K) opacities multiplied by $1\pm 0.02$;
low temperature opacities multiplied by $0.7 - 1.3$ (flat);
$\alpha-$capture abundances $[\alpha/\mathrm{Fe}] =
+0.2\,\mathrm{to}\,+0.7$ (flat); surface boundary conditions were
either gray or from \cite{kris}; color table from \cite{ryi} or
\cite{kurcol}; nuclear reaction rates mean values from \cite{newnuc},
with errors from \cite{mc2}.  The primordial helium abundance is
constrained to be in the narrow range $Y_p = 0.245\,\mathrm{to}\,
0.25$, motivated by recent advances in our ability to estimate this
quantity. Observations of deuterium in high redshift QSO absorption
systems, coupled with big bang nucleosynthesis (BBN) allow a reliable
estimate of the cosmic baryon fraction, $\Omega_Bh^2 =0.020 \pm
0.001$, where $h$ is the Hubble constant in units of 100 km/s/Mpc
\citep{burles}. The value determined by cosmic microwave background
experiments is $\Omega_bh^2 =0.022\pm 0.003$ \citep{cmb}. The
agreement between these two independent estimates is compelling and
allows us to interpret the bound on the baryon fraction, using BBN, in
terms of a new allowed range for $Y_p$.  Measurements in extragalactic
HII regions yield similar values of $Y_p = 0.245 \pm 0.004$
\citep{izotov}, although \cite{peimbert} find $Y_p = 0.238\pm 0.003$.
Using a value of $Y_p = 0.238$ results in SGB mass age which is 5\%
higher than that which is found using $Y_p = 0.245$.  However, given
the excellent agreement in determinations of $\Omega_bh^2$ from the
deuterium and CMB observations, we believe that it is very unlikely
that the true primordial helium abundance is as low as $Y = 0.238$.

The equation of state was not varied in the MC, as it is not thought
to be a significant source of error in stellar models.  To check this,
the SGB mass age of a $\mathrm{M} = 0.809\,\mathrm{M}_{\odot}$, $\feh
= -2.25$ star was calculated using stellar models calculated with the
OPAL equation of state \citep{rogers}, and a simple equation of state
which uses the Debye-H\"{u}ckel correction \citep{guenther}.  The two
sets of isochrones yielded ages which agreed with each other to within
$0.5\%$.
   
The only differences between the input parameters distributions in
this paper and in \cite{mc3} are for the nuclear reaction rates and
for the opacity.  As we discuss in the next section, the uncertainty
in the TO/SGB mass-age relation is dominated by the uncertainty in
opacity, leading us to critically re-examine the possible error
in modern opacity calculations.  There have been two recent studies
which have addressed the accuracy of opacity calculations for
conditions appropriate in the Sun.  \cite{rose} examined the
uncertainty in calculating the opacity at the solar core (a
temperature of $T = 1.6\times 10^6\,$K) by comparing the results of 7
different opacity codes.  \cite{rose} found a standard deviation of
5\% about the average.  \cite{opcompare} performed a detailed
comparison of the OPAL \citep{opal} and
LEDCOP\footnote{http://www.t4.lanl.gov/} \citep{ledcop} opacities
throughout the Sun.  They found that the OPAL and LEDCOP opacities
differ by $\sim 6\%$ at the base of the convection zone and by $\sim$
3\% at the solar core.

\centerline{\epsfclipon\epsfig{file=./f1.eps,height=8.0cm}}  
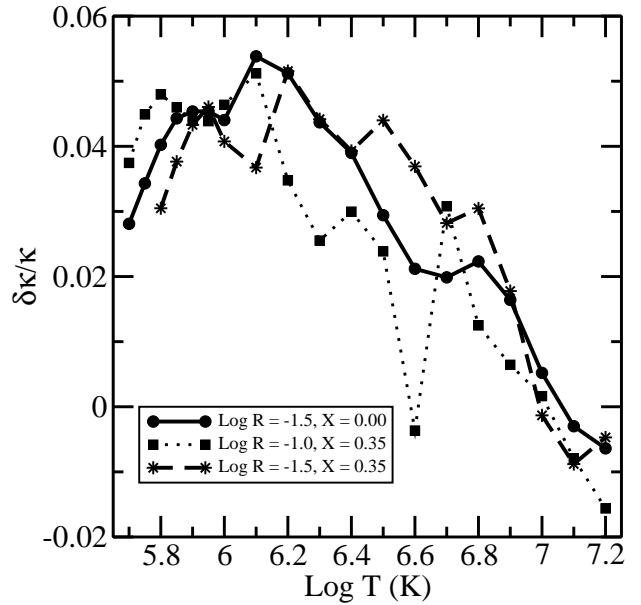
\hspace*{0.5cm} Fig. 1.---{\small Comparison between the OPAL and
LEDCOP opacities. The fractional difference in opacity plotted on the
$y$-axis is defined to be $\delta\kappa/\kappa = (\kappa_\mathrm{OPAL}
- \kappa_\mathrm{LEDCOP})/\kappa_\mathrm{OPAL}$. The differences in
the opacities have been calculated for a variety of hydrogen mass
fractions, $X$, and values of $\log R$ appropriate for the deep
interior of a metal-poor $\mathrm{M} = 0.80\,\mathrm{M}_{\odot}$ star
at the middle and end of it's main sequence lifetime.  In the stellar
model, the density/temperature parameter $\log R = \log (\rho/T_6^3)$
(where $T_6$ is the temperature in units of $10^6\,$K) is in the range
of $-1.5$ to $-1.0$ for the temperatures plotted.  }
\vspace*{0.5cm}

The conditions in GCs stars differ from the Sun in that there are
significantly fewer heavy elements.  This simplifies the opacity
calculations, and presumably the errors in low metallicity opacity
calculations will be smaller than in the solar case.  Figure 1 shows
differences between the OPAL and LEDCOP opacities for conditions
appropriate for a $\mathrm{M} = 0.80\,\mathrm{M}_{\odot}$ metal-poor
star at the middle and end of it's main sequence lifetime.  The OPAL and LEDCOP opacity calculations differ by $\sim
4\%$ at $\log T = 6.2$ and by $\sim 1\%$ around $\log T = 7$.

An independent estimate of the opacity for the conditions appropriate
for a the core of a main sequence, metal-poor star ($X = 0.35, Z =
0.0003$, $\log T = 7.2$, $\rho = 157\,\textrm{gm/cm}^3$, $\log R =
-1.4$) was calculated using the CASSANDRA opacity code
\citep{cassandra}.  For the same conditions, the CASSANDRA opacity was
0.5\% higher than the OPAL opacities and $0.4\%$ lower than the LEDCOP
opacities.  The OPAL and LEDCOP opacities for this data point were
determined via a simple linear interpolation (in $\log T$ and $\log
R$) in the public opacity tables.  
When the OPAL
opacity was calculated using the interpolation routines provided by
the OPAL group, the OPAL opacity was found to be 1.7\% lower than the
CASSANDRA opacity.  This suggests that the interpolation routine
introduce additional errors of order 1\% into the opacities used in
the stellar evolution code.
   
It is impressive that three difference opacity codes yield opacities
which agree to within 1\% for the conditions appropriate for the core
of a metal-poor star.  As stressed by \cite{opcompare}, the
true opacity could be different from the calculations, and Magee
(private communication) estimates a maximum uncertainty for these
conditions of 5\%.

\centerline{\epsfig{file=./f2.eps,height=8.0cm}}  
\hspace*{0.5cm} Fig. 2.---{\small 
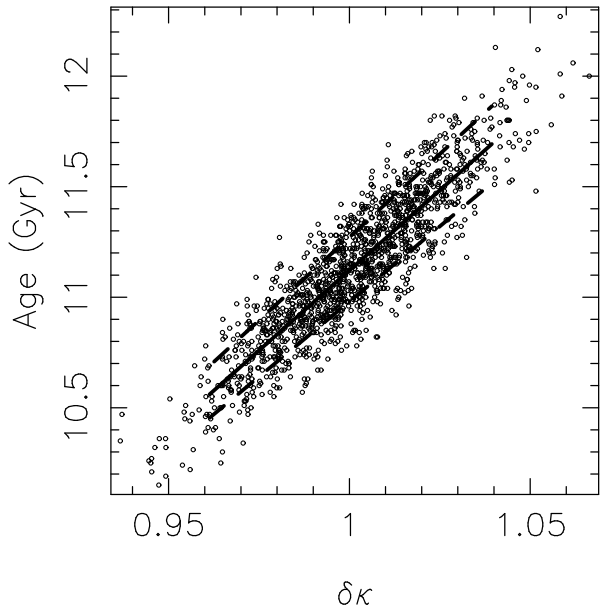
Dependence of the derived age of a SGB star on   
the high temperature ($T > 10^4$ K) opacities.  The $x$-axis, $\delta   
\kappa$ is the coefficient which is multiplying the opacities for $T   
\ge 10^6$ K. The solid line is the best fit to the median    
age and has the equation $t_9 = -3.38 + (14.50 \pm 0.37)\delta\kappa$,   
where $t_9$ is the age in Gyr.  The dotted lines are fits to the   
median $\pm 1\,\sigma$ points with the equations $t_9 = -2.62 +   
13.60\delta\kappa$ ($-1\,\sigma\,$) and $t_9 = -3.77 +   
15.04\delta\kappa$ ($+1\,\sigma\,$).   
}
\vspace*{0.5cm}

From Figure 1 it is clear that there is a systematic
difference between the OPAL and LEDCOP opacities, and that this
difference is a function of temperature.  At lower temperatures, the
OPAL opacities are always higher than the LEDCOP opacities.  To take
into account the systematic differences between the two opacity
calculations, the OPAL opacities (which are used in the stellar
evolution code) are multiplied by $0.98$ for $T\le 10^6\,$K and used
at their tabulated values for $T\ge 10^7\,$K.  Between $10^6\,$K and
$10^7\,$K, the multiplicative factor changes linearly.  From the 
opacity comparisons discussed previously, it is  clear that the
uncertainty in the opacity calculations increases with decreasing
temperatures.  As a result, we have taken the uncertainty in the
opacities to be Gaussian, with $\sigma = 4\%$ for $T\le 10^6\,$K, and
$\sigma = 2\%$ for $T\ge 10^7\,$K.  In between these two temperatures,
the Gaussian $\sigma$ changes linearly with temperature.
    
\section{Monte Carlo Results \label{monte}}   
In total 1500 different sets of input parameters were generated and
used to construct isochrones.  For each set of input parameters in the
MC, two isochrones were calculated with differing metallicities, $\feh
= -2.5$ and $\feh = -2.0$.  The set of 1500 MC isochrones was used to
determine the age of a SGB star, chosen to have properties similar to
OGLEGC-17, $\mathrm{M} = 0.809\,\mathrm{M}_{\odot}$, $\feh =
-2.25$. Furthermore, we fix the SGB star to be located 0.05 mag (in
\bv) from the TO. The resulting distribution of ages has a narrow
range with a Gaussian 1-$\sigma$ uncertainty of $\pm 2.9\%$.  This
confirms expectations that the SGB mass-age relation can be a robust
prediction of theoretical stellar evolution models \cite{pacz}.
    
The set of theoretical MC ages was analyzed to determine
which input parameters had a significant effect on the derived age.
The dominant source of error in deriving the age
of a star of fixed mass on the SGB using its mass are the 
high temperature ($T > 10^4\,$K) opacities. The
relationship between the 
\begin{center}
\begin{tabular}{lcc} 
\multicolumn{3}{c}{Table 1}\\
\multicolumn{3}{c}{Sensitivity of Age to Parameter Variations}\\
\hline
\multicolumn{1}{c}{Parameter}&
\multicolumn{1}{c}{$\delta$ Parameter} &
\multicolumn{1}{c}{$\delta$ Age (\%)}\\[2pt]
\hline
\hline
High Temperature Opacities & 2\% at $10^7\,$K & $+2.6$ \\
 
Helium Mass Fraction & 0.003 & $-1.4$ \\
 
[$\alpha$/Fe] & 0.2 dex & $+1.0$ \\
 
Helium Diffusion Coefficient & 30\% & $-1.0$ \\
\hline
\end{tabular}
\end{center}
 
\vspace*{0.5cm}  

\noindent
derived age and the opacity is shown in
Figure 2.  The solid line is the best fit to the median
age as a function of the opacity, and its slope implies that for every
1\% increase in the opacities at $10^7\,$K, the age will increase by
0.14 Gyr, or 1.3\%.  Given that the SGB age at a given
mass is essentially the main sequence lifetime of a given stellar
model, the relationship between age and opacity can be easily
understood given the equation of radiative transfer for a star which
implies $L \propto 1/\kappa$, where $L$ is the luminosity.  Hence, an
higher opacity leads to a decrease in the luminosity which in turn
results in an increase in the main sequence lifetime of a star of a
given mass.

The other input parameters in the MC had much smaller effects
on the derived age.  This can be readily seen in Figure~2, 
where the width of the age distribution at a given value of $\delta
\kappa$ gives an indication of the total uncertainty associated with
all of the other input parameters.  This width is about a factor of
two smaller than the range of ages in Figure 2.  If the error in the
opacity were zero, then from the $\pm 1\,\sigma$ fits shown in Figure
2 the total theoretical uncertainty in the derived ages would have a
Gaussian $\sigma = 1.3\%$.  This is somewhat more than a factor of two
smaller than the uncertainty found when including the uncertainty in
the opacities.  Besides opacity, the only parameters which lead to a
significant change in the derived age were the helium mass fraction
($Y$), the abundance of $\alpha-$capture elements, and the coefficient
of helium diffusion.  The effect that increasing each of these
parameters has on the age is summarized in Table 1.

\section{The Absolute Age of \wcen\ \label{age}}   
\cite{thompson} identified a number of detached eclipsing double line   
spectroscopic binaries in the GC \wcen\ and found that the primary in 
OGLEGC-17 was on the SGB,  ideally   
situated for an age determination.  \cite{kaluzny} report improved   
observations of OGLEGC-17 which yield a primary mass of $\mathrm{M} =   
0.809\pm 0.012\,\mathrm{M}_{\odot}$ and a metallicity of $\feh =   
-2.29\pm 0.15$.  It order to determine an accurate age 
for this star,   
one must determine is location relative to the metal-poor TO of   
\wcen.  An inspection of the color-magnitude diagram presented by   
\cite{thompson} leads us to conclude that OGLEGC-17 is located between   
0.03 and 0.07 mag redward (in \bv) of the metal-poor TO.  To   
determine the uncertainty in the age of OGLEGC-17, the following   
procedure was performed: (1) randomly pick an isochrone (out of our set   
of 1500 MC isochrones); (2) randomly pick a mass from the   
distribution $\mathrm{M} = 0.809\pm 0.012\,\mathrm{M}_{\odot}$; (3)   
randomly pick a metallicity using the distribution $\feh = -2.29 \pm   
0.15$; (4) randomly pick a location on the SGB by using a   
flat distribution which varied from 0.03 to 0.07 mag redward of   
the TO point and 
(5)~determine the age of OGLEGC-17 for this   
particular isochrone, mass, \feh\ and location on the SGB.   
This procedure was repeated 10,000 times.  The results are shown in
Figure  3.     
\centerline{\epsfclipon\epsfig{file=./f3.eps,height=8.0cm}}  
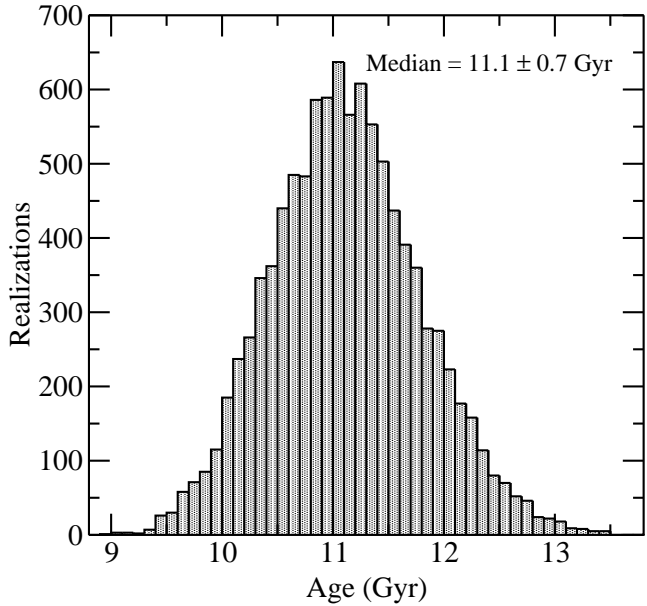
\hspace*{0.5cm} Fig. 3.---{\small Histogram of the derived age of the metal-poor SGB
in OGLEGC-17 in \wcen\ whose mass and metallicity were determined by    
\protect{\cite{kaluzny}}.  This histogram incorporates all known   
theoretical and observational errors, and reflects the total   
uncertainty in the age of OGLEGC-17. }

\vspace*{0.5cm}

\noindent   
   
The age of OGLEGC-17 determined in this way is $11.10\pm 0.67\,$Gyr; ie.\
the    total uncertainty in the age of this star is $\pm 6\%$.  The   
one-sided, 95\% lower limit to the age of OGLEGC-17 is 10.06 Gyr,   
while the one-sided, 95\% upper limit is 12.27 Gyr.  The derived   
uncertainty is dominated by the uncertainty of the mass determination.   
In the mass were known exactly, then the $1\,\sigma$ uncertainty in the   
derived age would be reduced to $\pm 3\%$.  Our derived age is fairly   
similar to that determined by \cite{kaluzny} who found $t = 11.8\pm   
0.6$ Gyr, assuming no error in the isochrones of \cite{girardi}.   

\section{Discussion \label{universe}}   
The age of OGLEGC-17 may be compared to our  estimate of the   
mean age of 17 metal-poor GCs which used the luminosity   
of the TO as an age indicator \citep{mc3}. For the same set of   
input parameters, we found a median age of 12.5 Gyr, and one sided 95\%   
confidence level ages of 10.2 Gyr and 15.9 Gyr.  The   
non-Gaussian distribution has a  lower $1\,\sigma$  age of 11.0   
Gyr, implying that age of OGLEGC-17 and the mean age of 17 metal-poor   
GCs agree at the $1\,\sigma$ level.  The one-sided 95\%   
confidence level lower limits to the two age determinations are quite   
similar (10.1 and 10.2 Gyr).  This supports our conclusion that the   
ages of the oldest stars and recent measurements of the Hubble   
constant require that the cosmic equation of state has $w \equiv \textrm{pressure/density} < -0.3$   
\citep{mc3}.   
   
The age of OGLEGC-17 was determined assuming that the error in the
mass determination was Gaussian.  As discussion of the error
in the mass determination of OGLEGC-17 has not been published it 
is not clear if this assumption is valid.  If it is, then age of
OGLEGC-17 is known much more accurately than the mean age of the
metal-poor GCs determined from their TO luminosity.  The upper limit
on the mean age (12.3 Gyr) is much smaller than that determined in the
GC study (15.9 Gyr).  The upper limit to the age of OGLEGC-17 may be
compared to the age of the universe determined from the cosmic
microwave background of $14.0\pm 0.5\,$Gyr \citep{knox}.  Their
$2\,\sigma$ lower limit of $13.0$ Gyr is 0.7 Gyr older than our upper
limit, implying at least 0.7 Gyr of galaxy evolution before OGLEGC-17
formed.  This corresponds to a redshift of globular cluster formation
of $z \la 7$ (cf.\ equation 1 in \cite{mc3}).  It is worth remarking
that when more old GC ages are constrained in this way, a comparison
strict upper limits one might derive on their ages with the Hubble age
may provide the strongest constraints on cosmological models with
exotic forms of dark energy such that $w= <-1$.
    
\acknowledgments 

We would like to thank Basil Crowley, Norman Magee
and Forest Rogers for their comments on the accuracy of opacity
calculations, Bohdan Paczy\'{n}ski for pointing out the \wcen\ results
to us, and Basil Crowley for supplying us with the CASSANDRA opacity
for a metal-poor mixture.

Research supported in part by a NSF CAREER grant 0094231  to BC and
a DOE grant to LMK.      BC is a Cottrell Scholar of the Research
Corporation.

\end{document}